%% file: paper.tex
\documentclass[prl,twocolumn,noshowpacs,noshowkeys,preprintnumbers,floatfix,
  nofootinbib,superscriptaddress]{revtex4-1}
\usepackage{amsfonts} 
\usepackage{amssymb} 
\usepackage{amsmath} 
\usepackage{subfigure}
\usepackage{graphicx} 
\usepackage{array} 
\usepackage{dcolumn} 
\usepackage{bm} 
\usepackage{latexsym} 
\usepackage{longtable} 
\usepackage{hyperref} 
\usepackage[usenames,dvipsnames]{xcolor}
\usepackage{mathrsfs}
\usepackage{comment}
\usepackage{rotating}
\usepackage{here}
\usepackage{float}
\usepackage{multirow}

\graphicspath{{./figs/}}

\allowdisplaybreaks

\input{macro.tex}

\begin{document}

\title{ Current status of $\varepsilon_K$ with lattice QCD inputs }
\author{Jon A. Bailey}
\affiliation{
  Lattice Gauge Theory Research Center, FPRD, and CTP, \\
  Department of Physics and Astronomy,
  Seoul National University, Seoul, 151-747, South Korea
}

\author{Yong-Chull Jang}
\affiliation{
  Lattice Gauge Theory Research Center, FPRD, and CTP, \\
  Department of Physics and Astronomy,
  Seoul National University, Seoul, 151-747, South Korea
}

\author{Weonjong Lee}
\email[E-mail: ]{wlee@snu.ac.kr}
\affiliation{
  Lattice Gauge Theory Research Center, FPRD, and CTP, \\
  Department of Physics and Astronomy,
  Seoul National University, Seoul, 151-747, South Korea
}

\author{Sungwoo Park}
\affiliation{
  Lattice Gauge Theory Research Center, FPRD, and CTP, \\
  Department of Physics and Astronomy,
  Seoul National University, Seoul, 151-747, South Korea
}

\collaboration{SWME Collaboration}
\date{\today}
\begin{abstract}
We present the Standard Model evaluation of the indirect CP violation
parameter $\varepsilon_K$ using inputs determined from lattice QCD
together with experiment: $|V_{us}|$, $|V_{cb}|$, $\xi_0$, and
$\hat{B}_K$.
We use the Wolfenstein parametrization ($|V_{cb}|$, $\lambda$,
$\bar{\rho}$, $\bar{\eta}$) for the CKM matrix elements.
For the central value, we take the angle-only fit of the UTfit
collaboration, and use $|V_{us}|$ from the $K_{\ell 3}$ and $K_{\mu
  2}$ decays as an independent input to fix $\lambda$.
For the error estimate, we use results of the global unitarity
triangle fits from the CKMfitter and UTfit collaborations.
We find that the Standard Model (SM) prediction of $\varepsilon_K$ with
exclusive $V_{cb}$ (lattice QCD results) is lower than the
experimental value by $3.6(2)\sigma$.
However, with inclusive $V_{cb}$ (results of the heavy quark
expansion), the tension between the SM prediction of $\varepsilon_K$
and its experimental value disappears.
\end{abstract}
\maketitle
%
%
%

%
CP violation in nature was first discovered in 1964 \cite{
  Christenson1964}.
The neutral kaon system has two kinds of CP violation: one is the
indirect CP violation due to CP-asymmetric impurity in the kaon
eigenstates in nature, and the other is the direct CP violation due to
the CP violating nature of the weak interaction \cite{
  AlaviHarati:1999xp, Fanti:1999nm}.
Here, we focus on the indirect CP violation, which is parametrized by
$\epsK$.
\begin{equation}
  \label{eq:epsK_def}
  \epsK 
  \equiv \frac{\mathcal{A}(K_L \to \pi\pi(I=0))} 
              {\mathcal{A}(K_S \to \pi\pi(I=0))} \,,
\end{equation}
where $K_L$ and $K_S$ are the neutral kaon eigenstates in nature, and
$I=0$ is the isospin of the final two-pion state.
In experiment \cite{Beringer2012:PhysRevD86.010001}, 
\begin{align}
  \label{eq:epsK_exp}
  \epsK &= (2.228 \pm 0.011) \times 10^{-3} 
  \times e^{i\phi_\eps} \,,\CL
  \phi_\eps &= 43.52 \pm 0.05 {}^\circ \,.
\end{align}
%

In the Standard Model (SM), the CP violation comes solely from a
single phase in the CKM matrix elements \cite{
  Kobayashi1973:ProgTheorPhys.49.652}.
The mixing of neutral kaons is allowed through the box
diagrams which describe the mass splitting $\Delta M_K$ and $\epsK$
\cite{ Winstein:1992sx, Buchalla:1995vs}.
In the SM, the master formula for $\epsK$ is 
\begin{align}
  \label{eq:epsK_SM_1}
  \epsK
  =& e^{i\theta} \sqrt{2}\sin{\theta} 
  \Big( C_{\eps} \hat{B}_{K} X_\text{SD} + \xi_{0} + \xi_\text{LD} \Big) \CL
   &+ \mathcal{O}(\omega\eps^\prime)
   + \mathcal{O}(\xi_0 \Gamma_2/\Gamma_1) \,,
\end{align}
where $C_{\eps}$ and $X_\text{SD}$ are defined as follows.
\begin{align}
  C_{\eps} 
  &= \frac{ G_{F}^{2} F_K^{2} m_{K^{0}} M_{W}^{2} }
                   { 6\sqrt{2} \pi^{2} \Delta M_{K} } \,,
\label{eq:C_eps} \\
  X_\text{SD} &= \bar{\eta}\lambda^2 \abs{V_{cb}}^2 
       \Bigg[ \abs{V_{cb}}^2 (1-\bar{\rho}) \eta_{tt} S_0(x_t) (1 + r) \CL
    &  \quad + \left(1-\frac{\lambda^4}{8}\right) 
       \left\{ \eta_{ct} S_0(x_c,x_t) - \eta_{cc} S_0(x_c) \right\} 
       \Bigg] \,,
\end{align}
where $S_0$'s are the Inami-Lim functions \cite{
  Inami1980:ProgTheorPhys.65.297}, and $x_i \equiv m_i^2/M_W^2$ with
$i = c,t$.
Here, $\displaystyle r = \{\eta_{cc} S_0(x_c) - 2\eta_{ct}
S_0(x_c,x_t)\}/\{\eta_{tt} S_0(x_t)\}$.
$\lambda$, $\bar{\rho}$, and $\bar{\eta}$ are the Wolfenstein
parameters of the CKM matrix elements \cite{Buchalla:1995vs}.
Here, we replace $A$ by $V_{cb}$, using the relation $\displaystyle
\abs{V_{cb}} = A \lambda^{2} + \mathcal{O}(\lambda^{8})$.
$\eta_{ij}$ with $i,j = c,t$ represents the QCD corrections to
the box diagrams.
$\xi_0 = \Im A_0 / \Re A_0$ represents the long distance effect from
the absorptive part, and $\xi_\text{LD}$ corresponds to the long
distance effect from the dispersive part \cite{ Bailey:2015tba}.
The correction terms $\mathcal{O}(\omega \varepsilon^\prime)$ and
$\mathcal{O}(\xi_0 \Gamma_2/\Gamma_1)$ are of order $10^{-7}$, which
we neglect in this paper.
The master formula of Eq.~\eqref{eq:epsK_SM_1} is essentially the
same as that of Ref.~\cite{ Buras2008:PhysRevD.78.033005}.
Details on how to derive Eq.~\eqref{eq:epsK_SM_1} from the SM are
given in our companion paper \cite{ Bailey:2015tba}.

In Eq.~\eqref{eq:epsK_SM_1}, the major contribution to $\epsK$ comes
from the $\BK$ term, and a minor contribution of about $-7\%$ comes
from the $\xi_0$ term.
The remaining contribution of $\xi_\text{LD}$ is about 2\%, coming from
the long distance effect on $\epsK$ \cite{
  Christ2012:PhysRevD.88.014508, Christ:2014qwa}.
In this paper, we neglect $\xi_\text{LD}$ without affecting our
conclusion.

The Wolfenstein parameters $\lambda,\bar{\rho},\bar{\eta}$ and $A$ can
be obtained from the global unitarity triangle (UT) fit.
Here, we use $\lambda,\bar{\rho},\bar{\eta}$ from the CKMfitter \cite{
  Charles:2004jd, Hocker:2001xe} and UTfit collaborations \cite{
  Bona:2005vz, Bona:2007vi}.
They are summarized in Table \ref{tab:wolfenstein}.

The parameters $\epsK$, $\BK$, and $V_{cb}$ are inputs to the global
UT fit.
Hence, the $\lambda,\bar{\rho},\bar{\eta}$ parameters extracted
from the global UT fit of the CKMfitter and UTfit groups contain
unwanted dependence on $\epsK$, $\BK$, and $V_{cb}$.
Therefore, in order to determine $\epsK$ self-consistently, we take
another input set from the angle-only-fit (AOF) in
Ref.~\cite{Bevan2013:npps241.89}.
Here the advantage is that the AOF does not use $\epsK, \hat{B}_K$,
and $V_{cb}$ as inputs to determine the UT apex parameters
$\bar{\rho}$ and $\bar{\eta}$.
The AOF gives the UT apex $(\bar{\rho}, \bar{\eta})$ but not
$\lambda$.
We can take $\lambda$ independently from the CKM matrix element
$V_{us}$, using the relation: $\abs{V_{us}} = \lambda +
\mathcal{O}(\lambda^7)$.
Here, the $K_{\ell3}$ and $K_{\mu2}$ decays are used to set $V_{us}$
\cite{Beringer2012:PhysRevD86.010001}.
\begin{table}[!th]
  \caption{Wolfenstein Parameters}
  \label{tab:wolfenstein}
  \renewcommand{\arraystretch}{1.2}
  \begin{ruledtabular}
  \begin{tabular}{clll}
  & CKMfitter & UTfit & AOF \\ \hline
  $\lambda$
  & $0.22535(65)$
  /\cite{Beringer2012:PhysRevD86.010001}
  & $0.22535(65)$
  /\cite{Beringer2012:PhysRevD86.010001}
  & $0.2252(9)$
  /\cite{Beringer2012:PhysRevD86.010001} \\ \hline
  $\bar{\rho}$
  & $\displaystyle 0.131^{+0.026}_{-0.013}$
  /\cite{Beringer2012:PhysRevD86.010001}
  & $0.136(18)$
  /\cite{Beringer2012:PhysRevD86.010001}
  & $0.130(27)$
  /\cite{Bevan2013:npps241.89} \\ \hline
  $\bar{\eta}$
  & $\displaystyle 0.345^{+0.013}_{-0.014}$
  /\cite{Beringer2012:PhysRevD86.010001}
  & $0.348(14)$
  /\cite{Beringer2012:PhysRevD86.010001}
  & $0.338(16)$
  /\cite{Bevan2013:npps241.89}
  \end{tabular}
  \end{ruledtabular}
\end{table}

In Table \ref{tab:Vcb}, we summarize the input values for $V_{cb}$.
In the inclusive channel, they use $ B \to X_c l \nu$, and $B
\to X_s \gamma $ decays.
They also use moments of outgoing lepton energy, hadron masses, and
photon energy and fit them to the theoretical expressions which come
from the operator product expansion (OPE).
They use perturbative expansion in the strong coupling constant
$\alpha_s$ and inverse heavy quark mass $\Lambda/m_b$.
For the $b$ quark mass, there are two popular schemes in the heavy
quark expansion: the kinetic scheme \cite{ Alberti:2014yda,
  Gambino2014:PhysRevD.89.014022} and the 1S scheme \cite{
  Beringer2012:PhysRevD86.010001}.
Here, we use the results of the kinetic scheme to calculate $\epsK$
since they are more recently updated and have somewhat larger errors.

For the exclusive channel for $V_{cb}$, we use the results of the
FNAL/MILC collaboration \cite{Bailey2014:PhysRevD.89.114504}.
They use lattice QCD to calculate the form factor $\mathcal{F}(w)$ of
the semi-leptonic decay $\bar{B}\to D^{\ast}\ell\bar{\nu}$ at zero
recoil ($w=1$).
The lattice measurements are done over the MILC $N_f=2+1$ asqtad gauge
ensembles \cite{Bazavov:2009bb}.
They use the Wilson clover action for the heavy quark with the
Fermilab interpretation \cite{El-khadra:PhysRevD.55.3933}.
They combine the lattice results for $\mathcal{F}(w)$ with the HFAG
average \cite{Amhis2012:HFAG} of experimental values of
$\mathcal{F}(1) \abs{\bar{\eta}_\text{EM}} \abs{V_{cb}}$ to extract
$V_{cb}$.
Here, $\abs{\bar{\eta}_\text{EM}}$ represents small enhancement factors
which come from electromagnetic corrections.
%
%
\begin{table}[!th]
  \caption{ Inclusive and exclusive $\abs{V_{cb}}$ in units of
    $10^{-3}$. Here, Kin.~represents the kinetic scheme in heavy quark
  expansion, and 1S the 1S scheme.}
  \label{tab:Vcb}
  \renewcommand{\arraystretch}{1.2}
  \begin{ruledtabular}
  \begin{tabular}{ccc}
  Inclusive (Kin.) & Inclusive (1S) & Exclusive \\ \hline
  $42.21(78)$
  /\cite{Alberti:2014yda}
  & $41.96(45)(07)$
  /\cite{Beringer2012:PhysRevD86.010001}
  & $39.04(49)(53)(19)$
  /\cite{Bailey2014:PhysRevD.89.114504}
  \end{tabular}
  \end{ruledtabular}
\end{table}

For the kaon bag parameter $\BK$, we use the FLAG average \cite{
  Aoki2013:hep-lat.1310.8555} and the SWME results \cite{
  Bae2014:prd.89.074504} which are summarized in Table \ref{tab:BK}.
FLAG combines several lattice results for $\BK$ with $N_f=2+1$ \cite{
  Bae2012:PhysRevLett.109.041601, Aoki2011:PhysRevD.84.014503,
  Aubin2010:PhysRevD.81.014507, Durr2011:PhysLettB.705.477} to obtain
the average.
FLAG uses the $\hat{B}_K$ result of the SWME collaboration \cite{
  Bae2012:PhysRevLett.109.041601}, which is not much different from
the most up-to-date value \cite{Bae2014:prd.89.074504} that we use in
this analysis.
The BMW calculation \cite{ Durr2011:PhysLettB.705.477} has the
smallest error, and it dominates the FLAG average.
The SWME result \cite{Bae2014:prd.89.074504} has a larger error, and
its value deviates most from the FLAG average.
%
%
\begin{table}[!th]
  \caption{$\BK$}
  \label{tab:BK}
  \renewcommand{\arraystretch}{1.2}
  \begin{ruledtabular}
  \begin{tabular}{cccc}
  & FLAG & SWME & \\ \hline
  & $0.7661(99)$
        /\cite{Aoki2013:hep-lat.1310.8555}
        & $0.7379(47)(365)$
        /\cite{Bae2014:prd.89.074504}
  &
        \end{tabular}
  \end{ruledtabular}
\end{table}

The RBC/UKQCD collaboration provides lattice results for $\mathrm{Im}
A_2$ and $\xi_0$ in Ref.~\cite{ Blum2011:PhysRevLett.108.141601}.
Here, we use their result of $\xi_0 = -1.63(19)(20)\times 10^{-4}$
obtained using the experimental value of $\eps^{\prime}/\eps$.

The factor $\eta_{tt}$ is given at next-to-leading order (NLO)
in Ref.~\cite{Buras2008:PhysRevD.78.033005}.
The factor $\eta_{ct}$ is given at next-to-next-to-leading order
(NNLO) in Ref.~\cite{Brod2010:prd.82.094026}.
The factor $\eta_{cc}$ is given at NNLO in Ref.~\cite{
  Brod2011:PhysRevLett.108.121801}.
In Ref.~\cite{ Buras2013:EurPhysJC.73.2560}, they claim that the error
is overestimated for the NNLO value of $\eta_{cc}$ given in
Ref.~\cite{ Brod2011:PhysRevLett.108.121801}.
Hence, in order to check the claim, we follow the renormalization
group (RG) evolution for $\eta_{cc}$ described in Ref.~\cite{
  Brod2011:PhysRevLett.108.121801} to produce the NNLO value of
$\eta_{cc}$.
The results are summarized in Table \ref{tab:cmp-etacc}.
In this table, note that the results are consistent with one another
within the systematic errors, but our $\eta_{cc}$ value is essentially
identical to that of Ref.~\cite{Buras2013:EurPhysJC.73.2560}.
Details of the SWME result are explained in Ref.~\cite{
  Bailey:2015tba}.
In this paper, we use the SWME result for $\eta_{cc}$ to obtain
$\epsK$.
\begin{table}[!th]
  \caption{Results of $\eta_{cc}$ at NNLO.}
  \label{tab:cmp-etacc}
  \renewcommand{\arraystretch}{1.2}
  \begin{ruledtabular}
  \begin{tabular}{ccc}
  Collaboration & Value & Ref. \\ \hline\hline
  Brod and Gorbahn & $1.86(76)$
  &\cite{Brod2011:PhysRevLett.108.121801} \\ \hline
  Buras and Girrbach & $1.70(21)$
  &\cite{Buras2013:EurPhysJC.73.2560} \\ \hline
  SWME & $1.72(27)$
  &\cite{ Bailey:2015tba}
  \end{tabular}
  \end{ruledtabular}
\end{table}

The input values for $\eta_{ij}$ that we use in this paper are
summarized in Table~\ref{tab:eta_ij}.
\begin{table}[!th]
  \caption{QCD corrections}
  \label{tab:eta_ij}
  \renewcommand{\arraystretch}{1.2}
  \begin{ruledtabular}
  \begin{tabular}{clc}
  Input & Value & Ref. \\ \hline\hline
  $\eta_{cc}$ & $1.72(27)$
  & \cite{ Bailey:2015tba} \\ \hline
  $\eta_{tt}$ & $0.5765(65)$
  &\cite{Buras2008:PhysRevD.78.033005} \\ \hline
  $\eta_{ct}$ & $0.496(47)$
  &\cite{Brod2010:prd.82.094026}
  \end{tabular}
  \end{ruledtabular}
\end{table}

The remaining input parameters are summarized in Table \ref{tab:rest}.
\begin{table}[!th]
  \caption{Other Input Parameters}
  \label{tab:rest}
  \renewcommand{\arraystretch}{1.2}
  \begin{ruledtabular}
  \begin{tabular}{clc}
  Input & Value & Ref. \\ \hline\hline
  $G_{F}$
        & $1.1663787(6) \times 10^{-5}$ GeV$^{-2}$
  &\cite{Beringer2012:PhysRevD86.010001} \\ \hline
  $M_{W}$
        & $80.385(15)$ GeV
  &\cite{Beringer2012:PhysRevD86.010001} \\ \hline
  $m_{c}(m_{c})$
        & $1.275(25)$ GeV
  &\cite{Beringer2012:PhysRevD86.010001} \\ \hline
  $m_{t}(m_{t})$
        & $163.3(2.7)$ GeV
  &\cite{Alekhin2012:plb.716.214} \\ \hline
  $\theta$
        & $43.52(5)^{\circ}$
        &\cite{Beringer2012:PhysRevD86.010001} \\ \hline
  $m_{K^{0}}$
        & $497.614(24)$ MeV
  &\cite{Beringer2012:PhysRevD86.010001} \\ \hline
  $\Delta M_{K}$
        & $3.484(6) \times 10^{-12}$ MeV
  &\cite{Beringer2012:PhysRevD86.010001} \\ \hline
  $F_K$
  & $156.1(8)$ MeV
  &\cite{Beringer2012:PhysRevD86.010001}
  \end{tabular}
  \end{ruledtabular}
\end{table}

Let us define $\epsK^\text{SM}$ as the theoretical evaluation of
$\abs{\eps_K}$ obtained using the master formula
Eq.~\eqref{eq:epsK_SM_1} directly from the SM.
We define $\epsK^\text{Exp}$ as the experimental value of
$\abs{\eps_K}$ given in Eq.~\eqref{eq:epsK_exp}.
We define $\Delta\epsK$ as the difference between
$\epsK^\text{Exp}$ and $\epsK^\text{SM}$:
\begin{equation}
  \Delta \epsK \equiv \epsK^\text{Exp} - \epsK^\text{SM} \,.
  \label{eq:DepsK}
\end{equation}
Here, we assume that the theoretical phase $\theta$ in
Eq.~\eqref{eq:epsK_SM_1} is equal to the experimental phase
$\phi_\eps$ in Eq.~\eqref{eq:epsK_exp} \cite{ Bailey:2015tba}.

\begin{figure}[!tbhp]
  \begin{center}
  \subfigure[$\;\epsK$ comparison (exclusive $V_{cb}$)]{%
    \includegraphics[width=0.40\textwidth]{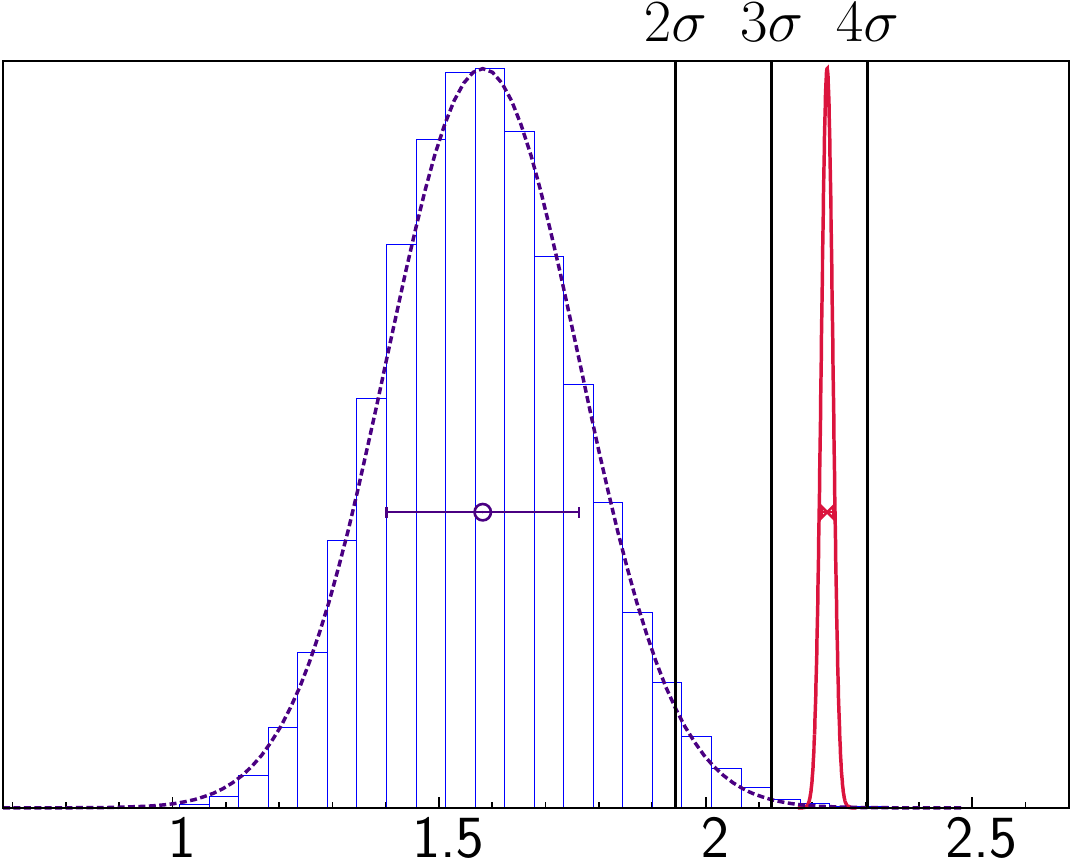}
    \label{sfig:excl.avg.AO}
  }
  \\
  \subfigure[$\;\epsK$ comparison (inclusive $V_{cb}$)]{%
    \includegraphics[width=0.40\textwidth]{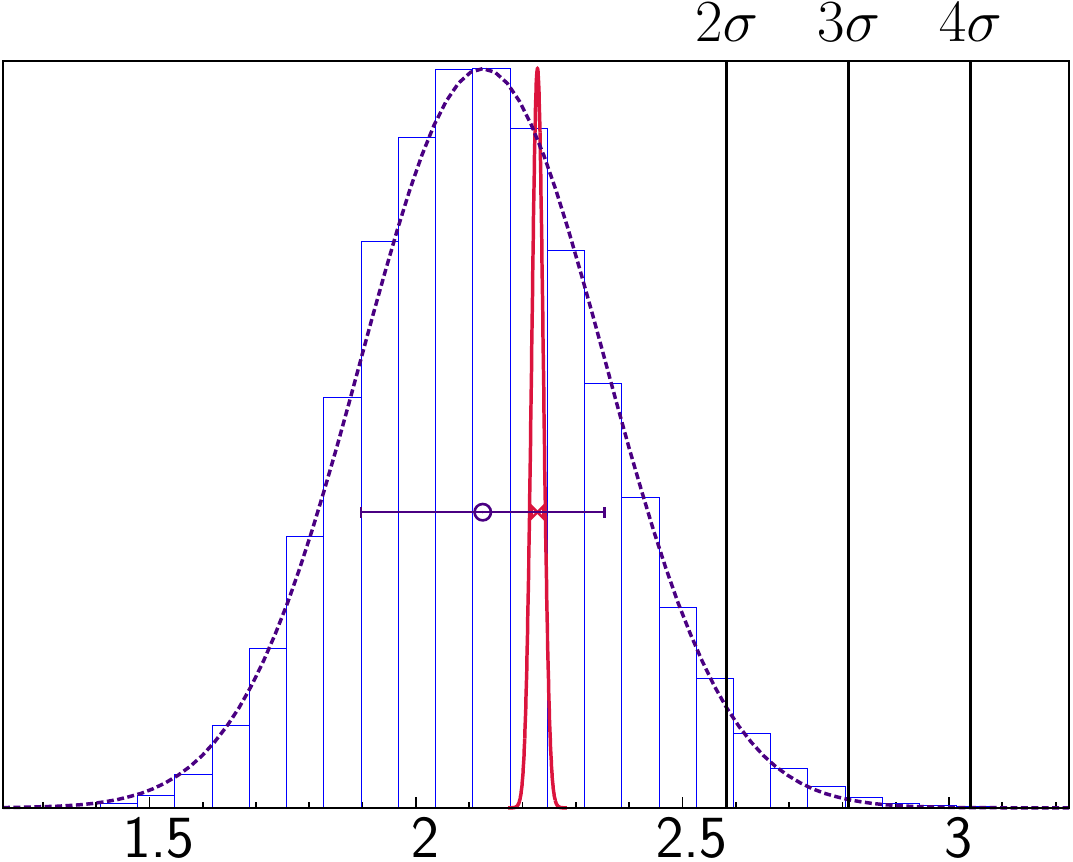}
    \label{sfig:incl.avg.AO}
  }
  \end{center}
  \caption{Probability distribution for $\epsK^\text{SM}$ (blue dotted
    line) and $\epsK^\text{Exp}$ (red solid line). The x-axis (y-axis)
    is the $\epsK$ value in the unit of $1.0 \times 10^{-3}$ (the
    probability distribution). Here, we use the FLAG $\BK$ and the AOF
    parameters. 
  }
  \label{fig:hstgwFLAG}
\end{figure}

In Table \ref{tbl:epsKwFLAG}, we present results for $\epsK^\text{SM}$
obtained using the FLAG average for $\BK$ together with $V_{cb}$ in
both inclusive and exclusive channels.
The corresponding probability distributions for $\epsK^\text{SM}$ and
$\epsK^\text{Exp}$ are presented in Fig.~\ref{fig:hstgwFLAG} for the
AOF case.
The corresponding results for $\Delta \epsK$ are presented in
Table \ref{tbl:DepsKwFLAG}.
%
%
%
\begin{table}[!tbh]
  \caption{ $\epsK^\text{SM}$ in the unit of $1.0 \times
    10^{-3}$. Here, we use the FLAG average for $\BK$ in Table
    \ref{tab:BK}.  The input methods of CKMfitter, UTfit, and AOF
    represent different inputs for the Wolfenstein parameters
    $\lambda, \bar{\rho}, \bar{\eta}$. }
  \label{tbl:epsKwFLAG}
  \renewcommand{\arraystretch}{1.2}
  \begin{ruledtabular}
  \begin{tabular}{ccc}
  Input Method & Inclusive $\Vcb$ & Exclusive $\Vcb$
  \\ \hline
  CKMfitter
  & $2.17(23)$ 
  & $1.62(18)$ 
  \\ \hline
  UTfit
  & $2.18(22)$ 
  & $1.63(18)$ 
  \\ \hline
  AOF
  & $2.13(23)$ 
  & $1.58(18)$ 
  \end{tabular}
  \end{ruledtabular}
\end{table}
\begin{table}[!tbh]
  \caption{$\Delta \epsK$. Here, we use $\epsK^\text{SM}$ from
    Table~\ref{tbl:epsKwFLAG}. We obtain $\sigma$ by combining
    the errors of $\epsK^\text{SM}$ and $\epsK^\text{Exp}$ in
    quadrature.}
  \label{tbl:DepsKwFLAG}
  \renewcommand{\arraystretch}{1.2}
  \begin{ruledtabular}
  \begin{tabular}{clc}
  Input Method & Inclusive $\Vcb$ & Exclusive $\Vcb$
  \\ \hline
  CKMfitter
  & $0.24\sigma$ 
  & $3.4\sigma$ 
  \\ \hline
  UTfit
  & $0.20\sigma$ 
  & $3.4\sigma$ 
  \\ \hline
  AOF
  & $0.44\sigma$ 
  & $3.6\sigma$ 
  \end{tabular}
  \end{ruledtabular}
\end{table}

From Table \ref{tbl:DepsKwFLAG}, we observe no tension between
$\epsK^\text{Exp}$ and $\epsK^\text{SM}$ with inclusive $V_{cb}$.

However, from Tables \ref{tbl:epsKwFLAG} and \ref{tbl:DepsKwFLAG}, we
find that $\epsK^\text{SM}$ with exclusive $V_{cb}$ is only 71\% of
$\epsK^\text{Exp}$.
For this case, with the most reliable input method (AOF),
$\Delta\epsK$ is $3.6\sigma$.
The largest contribution in this estimate of $\epsK^\text{SM}$ that we
have neglected is $\xi_\text{LD} \approx 2\%$.
Hence, the neglected contributions cannot explain the gap
$\Delta\epsK$ of 29\% with exclusive $V_{cb}$.
Hence, our final results for $\Delta\epsK$ are
\begin{align}
\Delta\epsK &= 3.6(2) \sigma & \qquad &\text{(exclusive $V_{cb}$)}
\label{eq:final-DepsK:ex} 
\\
\Delta\epsK &= 0.44(24) \sigma &\qquad &\text{(inclusive $V_{cb}$)}
\label{eq:final-DepsK:in}
\end{align}
where we take the AOF result as the central value and the systematic
error is obtained by taking the maximum difference among the input
methods in Table \ref{tbl:DepsKwFLAG}.

In the case of the FLAG $\BK$, the BMW result of $\BK$ \cite{
  Durr2011:PhysLettB.705.477} dominates the FLAG average, and the
gauge ensembles used for the BMW calculation are independent of those
used for the exclusive $V_{cb}$ \cite{ Bailey2014:PhysRevD.89.114504}.
Hence, we assume that we may neglect the correlation between the FLAG
$\BK$ and the exclusive $V_{cb}$.
However, the SWME calculation of $\BK$ in Ref.~\cite{
  Bae2014:prd.89.074504} shares the same MILC gauge ensembles with the
exclusive $V_{cb}$ determination in Ref.~\cite{
  Bailey2014:PhysRevD.89.114504}.
Hence, there exists a substantial correlation between the SWME $\BK$
and the exclusive $V_{cb}$.
We introduce $+50\%$ correlation and $-50\%$ anti-correlation between
the SWME $\BK$ and the exclusive $V_{cb}$ and take the maximum
deviation from the uncorrelated case as the systematic error due to
the unknown correlation between them.
Details of this analysis are explained in Ref.~\cite{ Bailey:2015tba}.
However, this analysis shows that the size of the ambiguity due to the
correlation between the SWME $\BK$ and the exclusive $V_{cb}$ is much
larger than the systematic error in $\Delta\epsK$ with the FLAG $\BK$.
Therefore, we use the results obtained with the SWME $\BK$ only to
cross-check those obtained with the FLAG $\BK$ \cite{ Bailey:2015tba}.

\begin{figure}[!tbhp]
  \centering
  \includegraphics[width=0.9\columnwidth]{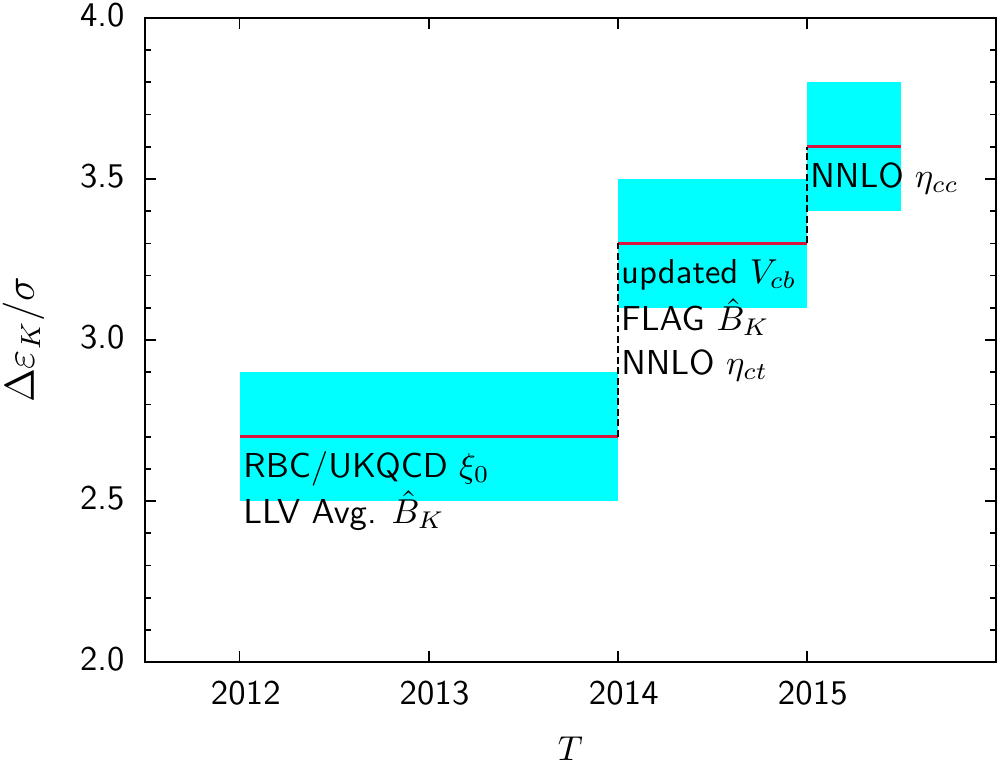}
  \caption{Recent history of $\Delta \epsK$ along with theoretical
    progress.}
  \label{fig:history}
\end{figure}
In Fig.~\ref{fig:history}, we present $\Delta \epsK/\sigma$ as a
function of time starting from 2012.
In 2012, the RBC/UKQCD collaboration reported $\xi_0$ in Ref.~\cite{
  Blum2011:PhysRevLett.108.141601}.
In addition to this, using the LLV average for $\BK$ \cite{
  Laiho2009:PhysRevD.81.034503}, we reported $\Delta\epsK =
2.7(2)\sigma$ in Ref.~\cite{ Jang2012:PoS.LAT2012.269} in 2012.
In 2014 FNAL/MILC reported an updated $V_{cb}$ from the exclusive
channel \cite{Bailey2014:PhysRevD.89.114504}.
Using the FLAG average for $\BK$ \cite{ Aoki2013:hep-lat.1310.8555}
and the NNLO value of $\eta_{ct}$ \cite{Brod2010:prd.82.094026}, we
reported the updated $\Delta\epsK = 3.3(2)\sigma$ in Ref.~\cite{
  Bailey:2014qda}.
In Ref.~\cite{Bailey:2015tba}, we investigate issues in the NNLO
calculation of $\eta_{cc}$ \cite{ Brod2011:PhysRevLett.108.121801,
  Buras2013:EurPhysJC.73.2560}, and in this paper we use the SWME
result in Table~\ref{tab:cmp-etacc} to report the updated $\Delta\epsK
= 3.6(2)\sigma$ in Eq.~\eqref{eq:final-DepsK:ex}.
%

In summary, we find that there is a substantial $3.6(2)\sigma$ tension
in $\epsK$ between the experiment and the SM theory with lattice QCD
inputs.
For this claim, we choose the angle-only fit (AOF), the exclusive
$V_{cb}$ (lattice QCD results), and the FLAG $\BK$ (lattice QCD
results) to determine the central value.
The systematic uncertainty is obtained by taking the maximum
deviation from the central value by choosing other input methods
from the global fits of the CKMfitter and UTfit.
We choose the AOF method to determine the central value because the
Wolfenstein parameters of AOF do not have unwanted correlation with
$\epsK$, $\BK$, and $|V_{cb}|$.
However, the tension disappears in the case of inclusive $V_{cb}$
(results of the heavy quark expansion based on the OPE).

In Table \ref{tbl:epsK-budget}, we present the error budget for
$\epsK^\text{SM}$ for the central value.
This is obtained using the error propagation method explained in
Ref.~\cite{ Bailey:2015tba}.
From this error budget, we observe that $V_{cb}$ dominates the error
in $\epsK^\text{SM}$.
Hence, it is essential to reduce the error of $V_{cb}$ as much as
possible (see also Refs.~\cite{ Buras:2014sba, Buras:2013ooa}).
To achieve this goal, there have been a lot of on-going efforts in the
lattice community \cite{ Oktay2008:PhysRevD.78.014504, Jang:LAT2013,
  Bailey:LAT2014, Jang:LAT2014, Atoui:2013zza, Monahan:2012dq}.
%
%
\begin{table}[tb!]
  \caption{Fractional error budget for $\epsK^\text{SM}$ obtained
    using the AOF method, the exclusive $\Vcb$, and the FLAG $\BK$.}
  \label{tbl:epsK-budget}
  \begin{ruledtabular}
  \begin{tabular}{ccc}
    source       & error (\%) & memo \\
    \hline
    $V_{cb}$     & 40.7        & FNAL/MILC \\
    $\bar{\eta}$ & 21.0        & AOF \\
    $\eta_{ct}$  & 17.2        & $c-t$ Box \\
    $\eta_{cc}$  &  7.3        & $c-c$ Box \\
    $\bar{\rho}$ &  4.7        & AOF \\
    $m_t$        &  2.5        & \\
    $\xi_0$      &  2.2        & RBC/UKQCD\\
    $\hat{B}_K$  &  1.6        & FLAG \\
    $m_c$        &  1.0        & \\
    $\vdots$     & $\vdots$    &
  \end{tabular}
  \end{ruledtabular}
\end{table}
%
%
%

It is true that there is an issue with the convergence of the
perturbative expansion of $\eta_{cc}$ \cite{
  Brod2011:PhysRevLett.108.121801}.
This could be resolved with lattice QCD calculations such as those
envisioned by the RBC/UKQCD collaboration \cite{
  Christ2012:PhysRevD.88.014508}.

We expect that our results for $\epsK$ would be consistent with those
from a global UT analysis, such as that in Ref.~\cite{
  Laiho2009:PhysRevD.81.034503}.
%


\begin{acknowledgments}
W.~Lee is supported by the Creative Research Initiatives program
(No.~2014001852) of the NRF grant funded by the Korean government
(MEST).
W.~Lee acknowledges support from the KISTI supercomputing
center through the strategic support program (No.~KSC-2013-G2-005).
J.A.B. is supported by the Basic Science Research Program of the
National Research Foundation of Korea (NRF) funded by the Ministry of
Education (No.~2014027937).
\end{acknowledgments}

\bibliographystyle{apsrev} 
\bibliography{ref} 

\end{document}

%% file: macro.tex
\providecommand{\abs}[1]{\lvert#1\rvert}

\renewcommand{\Re}{\mathrm{Re}}
\renewcommand{\Im}{\mathrm{Im}}

\providecommand{\CL}{\nonumber\\}

\definecolor{HLBlue}{HTML}{6599FF}
\definecolor{HLOrange}{HTML}{FF6600}

\newcommand{\BK}{\hat{B}_{K}}
\newcommand{\Vcb}{V_{cb}}

\newcommand{\eps}{\varepsilon}

\newcommand{\epsK}{\varepsilon_{K}}
